\documentclass{ws-procs975x65}

\begin{document}



\title{CAUSALITY AND SUPERLUMINAL FIELDS}

\author{JEAN-PHILIPPE BRUNETON}

\address{$\mathcal{G}\mathbb{R}\varepsilon\mathbb{C}\mathcal{O}$, Institut d'Astrophysique de Paris,
UMR 7095-CNRS, \\ Universit\'e Pierre et Marie Curie - Paris 6, 98
bis boulevard Arago F-75014, Paris, France \\
\email{bruneton@iap.fr}}


\begin{abstract}

The expression of causality depends on an underlying choice of
chronology. Since a chronology is provided by any Lorentzian
metric in relativistic theories, there are as many expressions of
causality as there are non-conformally related metrics over
spacetime. Although tempting, a definitive choice of a preferred
metric to which one may refer to is not satisfying. It would
indeed be in great conflict with the spirit of general covariance.
Moreover, a theory which appear to be non causal with respect to
(hereafter, w.r.t) this metric, may well be causal w.r.t another
metric. In a theory involving fields that propagate at different
speeds (e.g. due to some spontaneous breaking of Lorentz
invariance), spacetime is endowed with such a finite set of
non-conformally related metrics. In that case one must look for a
new notion of causality, such that 1. no particular metric is
favored and 2. there is an unique answer to the question : ``is
the theory causal?''. This new causality is unique and defined
w.r.t the metric drawing the wider cone in the tangent space of a
given point of the manifold. Moreover, which metric defines the
wider cone may depend on the location on spacetime. In that sense,
superluminal fields are generically causal, provided that some
other basic requirements are met.
\end{abstract}

\bodymatter

\section{Introduction}
Many authors argue that superluminal fields are not
causal\cite{Susskind,Arkani-Hamed,ListeDeRef,UtilisationPourContraindre}
(but see Refs.\cite{Daccordavecmoi}). This is not true, unless one
refers to an indefensible notion of causality. Indeed, as the
notion of causality evolves from Newtonian dynamics to Special
Relativity (SR), one must as well reconsider the notion of
causality from Special or General Relativity (GR), in which
spacetime is only endowed with the flat (resp. gravitational)
metric, to the case where it is endowed with a finite set of
Lorentzian metrics (notably then, if there are superluminal
fields).

In this short communication based on the more detailed
paper\cite{JPB06}, we thus look for an expression of causality in
such a multi-metric framework. The gravitational metric field is
denoted by $\mathbf{g}$, and $\mathcal{M}$ is a four-dimensional
differentiable manifold.

\section{Causality and chronology in field theories}
The analysis of the notion of causality leads in particular to the
following:
\newline
\textbf{Observation 1}: Since causes must precede effects,
causally connected events must be time-ordered. Causality thus
needs a notion of chronology to be expressed.
\newline
\textbf{Observation 2}: Any Lorentzian metric over $\mathcal{M}$
defines a local chronology (in the tangent space), through the
special relativistic notions of absolute future and past.

Gluing these two points together, we get the following
\newline
\textbf{Main point}: In relativistic field theories, there are
\textit{as many} notions of causality as there are non-conformally
related metrics over $\mathcal{M}$. These metrics $\mathbf{h}_i$
are the one along which the various fields $\psi_i$ propagate,
with velocities $c_i \neq c_j, \forall i \neq j$.

This plurality of the notion of causality is the crucial feature
of multi-metric theories. \textit{Indeed, it does not make any
sense to assert that a given theory is -or not- causal, if one
does not define to which metric (i.e. to which chronology) he
refers to}. A theory which appear to be non causal w.r.t some
metric may be causal w.r.t another metric.

To face this issue, one may be tempted to assume that there exists
a preferred metric field over $\mathcal{M}$. In other words, one
may fix a preferred chronology and its associated causal
structure. Most of the literature on causality and superluminal
fields is based -often implicitly- on this first approach. In
their famous textbook \cite{Hawking}, Hawking and Ellis recognize
explicitly that their notion of causality is defined w.r.t the
gravitational metric. This constitutes a ``\textit{postulate}
which sets the metric \textbf{g} apart from the other fields on
$\mathcal{M}$ and gives it its distinctive geometrical
character''\cite{Hawking} (p.60). As a consequence, fields that
propagate faster than gravitons are not causal. Thus, ``the null
cones of the matter equations [must] coincide or lie within the
null cone of the spacetime metric $\mathbf{g}$''\cite{Hawking}
(p.255).

Although such an attitude does not pose any problem when spacetime
is endowed with only one metric, as is the case of GR plus matter
fields that couple to $\mathbf{g}$, it becomes highly problematic
in the multi-metric case. First, indeed, there is no way to find
which metric should be favored, and which should not. Thus, by
invoking causality, different authors may find opposite
requirements on the theory\cite{JPB06}.

Second, let us consider two fields $\psi_i$ propagating along the
metrics $\mathbf{h}_i$ ($i=1,2$), such that $\psi_2$ travels
faster than $\psi_1$. Following the above reasonning, we can
define causality w.r.t the metric $\mathbf{h}_1$. Then two
observers that are spacelike related w.r.t $\mathbf{h}_1$ (and
hence, non time-ordered) but timelike related w.r.t $\mathbf{h}_2$
must be considered as causally disconnected, whereas they can
interact thanks to the field $\psi_2$. The only way to avoid so an
absurd conclusion is to define causality w.r.t to the metric that
defines the wider cone in the tangent space (see below).

Third, any choice of a preferred metric is equivalent to a choice
of preferred coordinates which, locally, diagonalize it. But the
existence of preferred coordinates, or equivalently, of preferred
rods and clocks\cite{JPB06}, is in great conflict with the whole
spirit of GR, namely diffeomorphism invariance; coordinates are
meaningless in GR.

The above attitude is thus irrelevant in the multi-metric case. As
an application, one should not invoke such a notion of causality
to put constraints on the theory (notably in order to fix various
signs), contrary to what is done in the
literature\cite{UtilisationPourContraindre}.

\section{An extended notion of causality and superluminal behaviors}

There is only one relevant notion of (extended) chronology that
does not refer to a given metric. This
consists\cite{MoffatClayton,JPB06} in defining the extended future
of a point $P$ as the union of the futures of $P$ defined by each
metric $\mathbf{h}_i$. The corresponding (extended) notion of
causality is thus in accordance with the notion of
interaction\footnote{The other possibility is to define an
extended future as the meet of each future of $P$. It would
however allow non-causally connected observers to interact, as in
the previous section.}. It is very permissive in the sense that,
by construction, any field theory is \textit{a priori} causal
provided that the various fields propagate along Lorentzian
metrics, so that the (extended) spacelike region is never empty.
Moreover, interactions cannot threaten this causal behavior,
since, by construction, the extended future and past are defined
at \textit{each} point of $\mathcal{M}$. Which metric defines the
wider cone may thus depend on the location on spacetime. In
particular, superluminal fields are {\it a priori} causal.

Of course, this construction is not sufficient. Causality also
requires, first, that the whole theory has an initial value
formulation. This is generically the case if the field equations
form a quasilinear, diagonal and second order hyperbolic
system\cite{Hawking,JPB06}. Beware however that initial data must
be assigned on hypersurfaces that are spacelike in the extended
sense, that is spacelike w.r.t to {\it all} metrics
$\mathbf{h}_i$. All the difficulties in the Cauchy problem of
superluminal fields found in the literature arise from an
irrelevant choice of initial data
surfaces\cite{Susskind,Arkani-Hamed}. Second, a {\it local}
chronology is not enough. We must have at hand a global chronology
over spacetime, in order to prevent, e.g. the existence of closed
timelike curves. In the multi-metric case, we shall also require
that our extended chronology is a global one, that is that no
closed extended-timelike curves exist.

It has been shown\cite{Arkani-Hamed} that a particular
superluminal scalar field may suffer from such a global pathology.
This is however not enough to kill this theory, for the very
reason that GR itself may suffer from such causal anomalies.
Therefore, difficulties at a global level {\it do not} signal an
intrinsic disease of superluminal fields. Rather, they originate
from the fact that the global topology of the Universe is not
imposed by local field equations. It is therefore necessary to
\textit{assume} that spacetime does not involve any closed
(extended) timelike curves to ensure causality.

\vfill

\end{document}